# Deep Random based Key Exchange protocol resisting unlimited MITM

Thibault de Valroger (*)

Abstract

We present a protocol enabling two legitimate partners sharing an initial secret to mutually authenticate and to exchange an encryption session key. The opponent is an active Man In The Middle (MITM) with unlimited calculation and storage capacities. The resistance to unlimitedly powered MITM is obtained through the combined use of Deep Random secrecy, formerly introduced [9] and proved as unconditionally secure against passive opponent for key exchange, and universal hash techniques. We prove the resistance to MITM interception attacks, and show that (i) upon successful completion, the protocol leaks no residual information about the current value of the shared secret to the opponent, and (ii) that any unsuccessful completion is detectable by the legitimate partners. We also discuss implementation techniques.

## I. Introduction and summary of former work

Modern cryptography mostly relies on mathematical problems commonly trusted as very difficult to solve, such as large integer factorization or discrete logarithm, belonging to complexity theory. No certainty exists on the actual difficulty of those problems. Some other methods, rather based on information theory, have been developed since early 90's. Those methods relies on hypothesis about the opponent (such as « memory bounded » adversary [6]) or about the communication channel (such as « independent noisy channels » [5]) ; unfortunately, if their perfect secrecy have been proven under given hypothesis, none of those hypothesis are easy to ensure in practice. At last, some other methods based on physical theories like quantum indetermination [3] or chaos generation have been described and experimented, but they are complex to implement, and, again, relies on solid but not proven and still partly understood theories.

Considering this theoretically unsatisfying situation, we have proposed in [9] to explore a new path, where proven information theoretic security can be reached, without assuming any limitation about the opponent, who is supposed to have unlimited calculation and storage power, nor about the communication channel, that is supposed to be perfectly public, accessible and equivalent for any playing party (legitimate partners and opponents). Furthermore, while we were only considering passive unlimited opponents in [9], we consider in this work active unlimited MITM opponents.

*(*) See contact and information about the author at last page*

In our model of security, the legitimate partners of the protocol are using Deep Random generation to generate their shared encryption key, and the behavior of the opponent, when inferring from public information, is governed by Deep Random assumption, that we introduce. The legitimate partners have an initial shared authentication secret, that is used only for authentication purpose, not for generating the shared encryption key.

Our protocol is original because it resists to opponents that are both unlimited and active MITM. A scheme like Wigner-Carter Authentication [11] is secure against unlimited opponent but not in a MITM scenario. WCA can however be secure against MITM in the specific case of QKD because of the non-cloning theorem. Public Key schemes can be resistant to MITM opponent but not if they are unlimited. Methods like tamper detection by latency examination [12] are not robust enough and not resistant to unlimited opponents.

## Back on the Deep random assumption

We have introduced in [9] the Deep Random assumption, based on Prior Probability theory as developed by Jaynes [7]. Deep Random assumption is an objective principle to assign probability, compatible with the symmetry principle proposed by Jaynes [7].

Before presenting the Deep Random assumption, it is needed to introduce Prior probability theory.

If we denote $\mathfrak{I}_<$ the set of all prior information available to observer regarding the probability distribution of a certain random variable $X$ ('prior' meaning before having observed any experiment of that variable), and $\mathfrak{I}_>$ any public information available regarding an experiment of $X$, it is then possible to define the set of possible distributions that are compatible with the information $\mathfrak{I} \triangleq \mathfrak{I}_< \cup \mathfrak{I}_>$ regarding an experiment of $X$; we denote this set of possible distributions as:

$$D_{\mathfrak{I}}$$

The goal of Prior probability theory is to provide tools enabling to make rigorous inference reasoning in a context of partial knowledge of probability distributions. A key idea for that purpose is to consider groups of transformation, applicable to the sample space of a random variable $X$, that do not change the global perception of the observer. In other words, for any transformation $\tau$ of such group, the observer has no information enabling him to privilege $\varphi_{\mathfrak{I}}(v) \triangleq P(X = v|\mathfrak{I})$ rather than $\varphi_{\mathfrak{I}} \circ \tau(v) = P(X = \tau(v)|\mathfrak{I})$ as the actual conditional distribution. This idea has been developed by Jaynes [7].

We will consider only finite groups of transformation, because one manipulates only discrete and bounded objects in digital communications. We define the acceptable groups $G$ as the ones fulfilling the 2 conditions below:

$(C1)$ Stability - For any distribution $\varphi_{\mathfrak{I}} \in D_{\mathfrak{I}}$, and for any transformation $\tau \in G$, then $\varphi_{\mathfrak{I}} \circ \tau \in D_{\mathfrak{I}}$

$(C2)$ Convexity - Any distribution that is invariant by action of $G$ does belong to $D_{\mathfrak{I}}$

It can be noted that the set of distributions that are invariant by action of $G$ is exactly:

$$R_{\mathfrak{I}}(G) \triangleq \left\{ \frac{1}{|G|} \sum_{\tau \in G} \varphi_{\mathfrak{I}} \circ \tau \,|\forall \varphi_{\mathfrak{I}} \in D_{\mathfrak{I}} \right\}$$

For any group $G$ of transformations applying on the sample space $F$, we denote by $\Omega_{\mathfrak{I}}(G)$ the set of all possible conditional expectations when the distribution of $X$ courses $R_{\mathfrak{I}}(G)$. In other words:

$$\Omega_{\mathfrak{I}}(G) \triangleq \{Z(\mathfrak{I}) \triangleq E[X|\mathfrak{I}] | \forall \varphi_{\mathfrak{I}} \in R_{\mathfrak{I}}(G)\}$$

Or also:

$$\Omega_{\mathfrak{I}}(G) = \left\{Z(\mathfrak{I}) = \int_F v\varphi_{\mathfrak{I}}(v)dv \,|\forall \varphi_{\mathfrak{I}} \in R_{\mathfrak{I}}(G)\right\}$$

The **Deep Random assumption** prescribes that, if $G \in \Gamma_{\mathfrak{I}}$, the strategy $Z_\xi$ of the opponent observer $\xi$, in order to estimate $X$ from the public information $\mathfrak{I}$, should be chosen by the opponent observer $\xi$ within the restricted set of strategies:

$$\boldsymbol{Z_\xi \in \Omega_{\mathfrak{I}}(G)} \qquad (\boldsymbol{A})$$

The Deep Random assumption can thus be seen as a way to restrict the possibilities of $\xi$ to choose his strategy in order estimate the private information $X$ from his knowledge of the public information $\mathfrak{I}$. It is a fully reasonable assumption because the assigned prior distribution should remain stable by action of a transformation that let the distribution uncertainty unchanged.

$(A)$ suggests of course that $Z_\xi$ should eventually be picked in $\bigcap_{G\in\Gamma_{\mathfrak{I}}} \Omega_{\mathfrak{I}}(G)$, but it is enough for our purpose to find at least one group of transformation with which one can apply efficiently the Deep Random assumption to the a protocol in order to measure an advantage distilled by the legitimate partners compared to the opponent.

**Back on the presentation of protocol $\mathcal{P}$ (introduced in [9])**

The following protocol has been presented in [9]. In order to shortly remind the notations, the sample space of the distribution of the private information (for $A$ or $B$) is $[0,1]^n$. Considering $x = (x_1, \dots, x_n)$ and $y = (y_1, \dots, y_n)$ some parameter vectors in $[0,1]^n$ and $i = (i_1, \dots, i_n)$ and $j = (j_1, \dots, j_n)$ some Bernoulli experiment vectors in $\{0,1\}^n$, we denote :

$x.y$ (resp. $i.j$) the scalar product of $x$ and $y$ (resp. $i$ and $j$)

$$|x| \triangleq \sum_{s=1}^n x_s \, ; \, |i| \triangleq \sum_{s=1}^n i_s$$

$\forall \sigma \in \mathfrak{S}_n$, $\sigma(x)$ represents $(x_{\sigma(1)}, \dots, x_{\sigma(n)})$

$\frac{x}{k}$ represents $\left(\frac{x_1}{k}, \dots, \frac{x_n}{k}\right)$ for $k \in \mathbb{R}_+^*$

In that protocol, besides being hidden to any third party (opponent or partner), the probability distribution used by each legitimate partner also needs to have specific properties in order to prevent the opponent to efficiently evaluate $V_A$ by using internal symmetry of the distribution.

Those specific properties are:

(i) Each probability distribution $\Phi$ (for $A$ or $B$) must be « far » from its symmetric projection $\overline{\Phi}(x) \triangleq \frac{1}{n!}\sum_{\sigma\in\mathfrak{S}_n} \Phi \circ \sigma(x)$

(ii) At least one of the distribution (of $A$ or $B$) must avoid to have brutal variations (Dirac)

The technical details explaining those constraints are presented in [9]. The set of compliant distributions is denoted $\zeta(\alpha)$ where $\alpha$ is a parameter that measures the « remoteness » of a distribution from its symmetric projection.

For such a distribution $\Phi$, a tidying permutation, denoted $\sigma_\Phi$, is a specific permutation that enables to give a canonical form $\Phi \circ \sigma_\Phi$ of $\Phi$, such form being useful to « synchronize » two distributions by transitivity. Again, technical details are given in [9]. One can just say here that it is linked to the quadratic matrix whose coefficient is $M_\Phi(u,v) \triangleq \int_{[0,1]^n} x_u x_v \Phi(x) dx$, by minimizing

$$\sum_{u,v \in I_0 \times \bar{I}_0} M_{\Phi \circ \sigma_\Phi}(u,v) = \min_{\sigma \in \mathfrak{S}_n} \left( \sum_{u,v \in I_0 \times \bar{I}_0} M_{\Phi \circ \sigma}(u,v) \right)$$

where $I_0 \triangleq \{1, \dots, {}^{n}/_{2}\}$.

Here are the steps of the proposed protocol:

$A$ and $B$ are the legitimate partners. The steps of the protocol $\mathcal{P}(\alpha, n, k, L)$ are the followings:

*Step 1 – Deep Random Generation: A and B pick independently the respective probability distributions* $\Phi$ *and* $\Phi' \in \zeta(\alpha)$*, so that* $\Phi$ *(resp.* $\Phi'$*) is secret (under Deep Random assumption) for any observer other than A (resp. B) beholding all the published information. A draws the parameter vector* $x \in [0,1]^n$ *from* $\Phi$*. B draws the parameter vector* $y \in [0,1]^n$ *from* $\Phi'$*.*

*Step 2 – Degradation: A generates a Bernoulli experiment vectors* $i \in \{0,1\}^n$ *from the parameter vector* $\frac{x}{k}$*. A publishes i. B generates a Bernoulli experiment vectors* $j \in \{0,1\}^n$ *from the parameter vector* $\frac{y}{k}$*. B publishes j.*

*Step 3 – Dispersion: A and B also pick respectively a second probability distribution* $\Psi$ *and* $\Psi' \in \zeta(\alpha)$ *such that it is also secret (under Deep Random assumption) for any observer other than A (resp. B).* $\Psi$ *is selected also such that* $\int_{|x| \in [k|i| - \sqrt{n}, k|i| + \sqrt{n}]} \Psi(x) dx \geq \frac{1}{2\sqrt{n}}$ *in order to ensure that* $|i|$ *is not an unlikely value for* $\approx \left|\frac{x}{k}\right|$ *(same for* $\Psi'$ *by replacing x by y and i by j).* $\Psi$ *(resp.* $\Psi'$*) is used to scramble the publication of the tidying permutation of A (resp. B). A (resp. B) computes a permutation* $\sigma_d[i]$ *(resp.* $\sigma'_d[j]$*) representing the reverse of the most likely tidying permutation on* $\Psi$ *(resp.* $\Psi'$*) to produce i (resp. j). In other words, with i,* $\sigma_d[i]$ *realizes :*

$$\max_{\sigma \in \mathfrak{S}_n} \int_x P(i|x) \Psi \circ \sigma_\Psi \circ \sigma^{-1}(x) dx$$

*Then A (resp. B) draws a boolean* $b \in \{0,1\}$ *(resp.* $b'$*) and publishes in a random order* $(\mu_1, \mu_2) = t^b(\sigma_d[i], \sigma_\Phi)$*, (resp.* $(\mu'_1, \mu'_2) = t^{b'}(\sigma'_d[j], \sigma_{\Phi'})$*) where t represents the transposition of elements in a couple.*

*Step 4 – Synchronization: A (resp. B) chooses randomly* $\sigma_A$ *(resp.* $\sigma_B$*) among* $(\mu'_1, \mu'_2)$ *(resp.* $(\mu_1, \mu_2)$*).*

*Step 5 – Decorrelation: $A$ computes $V_A = \frac{\sigma_\Phi{}^{-1}(x).\sigma_{\mathrm{A}}{}^{-1}(j)}{n}$, $B$ computes $V_B = \frac{\sigma_{\mathrm{B}}{}^{-1}(i).\sigma_{\Phi'}{}^{-1}(y)}{n}$. $V_A$ and $V_B$ are then transformed respectively by $A$ and $B$ in binary output thanks to a sampling method described hereafter. At this stage the protocol can then be seen as a broadcast model with 2 Binary Symmetric Channels (BSC), one between $A$ and $B$ and one between $A$ and $\xi$ who computes a certain $V_\xi$, called $\xi$'s strategy, that is to be transformed in binary output by the same sampling method than for $A$ and $B$. It is shown in Theorem 1 of [9] that those 2 BSC are partially independent, which enable to create Advantage Distillation as shown in [5].*

*Step 5' – Advantage Distillation: by applying error correcting techniques with code words of length $L$ between $A$ and $B$, as introduced in [5], we show in Theorem 1 of [9] that we can then create advantage for $B$ compared to $\xi$ in the error rates of the binary flows resulting from the error correcting code.*

*Step 6: classical Information Reconciliation and Privacy Amplification (IRPA) techniques then lead to get accuracy as close as desired from perfection between estimations of legitimate partners, and knowledge as close as desired from zero by any unlimitedly powered opponent, as shown in [4].*

The choices of the parameters $(\alpha, n, k, L)$ are theoretically discussed in proof of main Theorem in [9]. They are set to make steps 5, 5' and 6 possible.

The Degradation transformations $x \longmapsto \frac{x}{k}$ and $y \longmapsto \frac{y}{k}$ with $k > 1$ at step 2 are the ones that prevent the use of direct inference by the opponent, and of course, the Deep Random Generation at step 1 prevents the use of Bayesian inference based on the knowledge of the probability distribution. The synchronization step 4 is designed to overcome the independence between the choices of the distributions of $A$ and $B$, and needs that the distributions to have special properties ($\in \zeta(\alpha)$) in order to efficiently play their role. It is efficient in $1/4$ of cases (when $B$ picks $\sigma_B = \sigma_\Phi$ and $A$ picks $\sigma_A = \sigma_{\Phi'}$, which we will call favorable cases). And to prevent $\xi$ from gaining knowledge of $\sigma_\Phi$, Dispersion step 3 mixes $\sigma_\Phi$ within $(\mu_1, \mu_2)$ with another permutation $\sigma_d[i]$ (and $\sigma_{\Phi'}$ within $(\mu'_1, \mu'_2)$ with another permutation $\sigma'_d[j]$) that (1) is undistinguishable from $\sigma_\Phi$ knowing $i$, and (2) manages to make the estimation of $\xi$ unefficient as shown in [9]. We denote the following set of strategies (invariant by transposition of $(\mu_1, \mu_2)$ or $(\mu'_1, \mu'_2)$):

$$\Omega'_\# \triangleq \left\{\omega(i, j, (\mu_1, \mu_2), (\mu'_1, \mu'_2)) \,\middle|\, \forall b, b' \in \{0,1\} : \omega\left(i, j, \tau^b(\mu_1, \mu_2), \tau^{b'}\left(\mu'_1, \mu'_2\right)\right) = \omega(i, j, (\mu_1, \mu_2), (\mu'_1, \mu'_2))\right\}$$

Because of Deep Random assumption $(A)$ over the group $\{\mathrm{Id}, t\}$ applied to the distribution of $(\mu_1, \mu_2)$ and $(\mu'_1, \mu'_2)$, the strategy of the opponent can thus be restricted to $V_\xi \in \Omega'_\#$.

$i$ is entirely determined by $|i|$ and a permutation, which explains the constraint and transformation applied on $\Psi$ in step 3 to make $\sigma_\Phi$ and $\sigma_d[i]$ indisguishable knowing $i$ (same with $\sigma'_d[j]$, $\sigma_{\Phi'}$, and $j$).

The synchronization step has a cost when considering the favorable cases: $\xi$ knows that $\Phi$ and $\Phi'$ are synchronized in favorable cases, which means in other words that $\xi$ knows that an optimal (or quasi optimal) permutation is applied to $\Phi'$. This also means that in favorable cases, all happen like if when A picks $\Phi \circ \sigma$ instead of $\Phi$, the result of the synchronization is that $B$ uses $\Phi' \circ \sigma$ instead of $\Phi'$.

Starting from the most general strategy $\omega \in \Omega'_{\#}$ for $\xi$, we also consider the following additional restrictions applicable to the favorable cases:

- Restriction to the strategies of the form $\omega(i,j)$, because $(\sigma_d[i], \sigma'_d[j])$ depends only on $(i,j)$ and not on $\Phi$ neither $\Phi'$,
- And then restriction to the set of strategies such that $\omega_{i,j} = \omega_{\sigma(i),\sigma(j)}, \forall \sigma \in \mathfrak{S}_n$, in other words strategies invariant by common permutation on $i,j$.

which leads to define the more restricted set of strategies:

$$\Omega_{\#}(G,\mathcal{P}) = \left\{\omega \in [0,1]^{2^{2n}} | \omega(\sigma(i),\sigma(j),\mu_1,\mu_2,\mu'_1,\mu'_2) = \omega(\sigma(i),\sigma(j)) = \omega(i,j)\ ,\forall \sigma \in \mathfrak{S}_n\right\}$$

The step 5 is called Decorrelation because at this step, thanks to the Deep Random Assumption, we have managed to create a protocol that can be equivalently modelized by a broadcast communication over 2 partially independent (not fully correlated) BSC, as shown in the main Theorem in [9], and also that consequently, it is possible to apply error correcting techniques to create Advantage Distillation as established in [5].

We can use the following very basic technique to transform $V_A$ and $V_B$ in an intermediate binary flow as introduced in step 5: we can typically sample a value in $[0,1]$ like $V_A$ or $V_B$ with a gauge being a multiple of the variance $E\left[\left(V_{A|\sigma_A=\sigma_{\Phi'}} - V_{B|\sigma_B=\sigma_\Phi}\right)^2\right]^{1/2} = O\left(\frac{1}{\sqrt{nk}}\right)$. The multiplicative factor $K$ is chosen such that :

$$\frac{1}{\sqrt{nk}} \ll \frac{K}{\sqrt{nk}} \ll \frac{1}{\sqrt{n}}$$

and therefore, each experiment of the protocol can lead respectively $A$ and $B$ to distill an intermediate digit defined by :

$$\widetilde{e_A} = \left\lfloor \frac{V_A\sqrt{nk}}{K} \right\rfloor \bmod 2,\ \widetilde{e_B} = \left\lfloor \frac{V_B\sqrt{nk}}{K} \right\rfloor \bmod 2$$

Regarding the legitimate partners, when $B$ picks $\sigma_B = \sigma_\Phi$ and $A$ picks $\sigma_A = \sigma_{\Phi'}$, the choice of $\sigma_A$ and $\sigma_B$ remain independent from $i,j$, so that $i$ and $j$ remain draws of independent Bernoulli random variables, then allowing to apply Chernoff-style bounds for the legitimate partners. When $B$ picks $\sigma_B = \sigma_d[i]$ or $A$ picks $\sigma_A = \sigma'_d[j]$, this is no longer true and $V_B$ or $V_A$ become erratic, which will lead to error detection by error correcting code at step 5'.

The heuristic table analysis of the protocol is then the following:

| | $B$ picks $\sigma_B = \sigma_\Phi$ among $(\mu_1, \mu_2)$ | $B$ picks $\sigma_B = \sigma_d[i]$ among $(\mu_1, \mu_2)$ |
|---|---|---|
| $A$ picks $\sigma_A = \sigma_{\Phi'}$ among $(\mu'_1, \mu'_2)$ | $A$ and $B$ respective estimations are close in ~100% of cases, and thus both obtain accurate estimation of the combined shared secret in ~100% of cases.<br><br>$\xi$ cannot make accurate estimation of the combined shared secret in at least ~25% of cases (if $\xi$ tries to have a strategy depending on $(\mu_1, \mu_2, \mu'_1, \mu'_2)$, then $(\sigma_d[i], \sigma'_d[j])$ is indistinguishable from $(\sigma_\Phi, \sigma_{\Phi'})$ and is thus picked by $\xi$ in 25% of cases. | $A$ and $B$ respective estimations are not close which leads to error detection and finally discarding. |
| $A$ picks $\sigma_A = \sigma'_d[j]$ among $(\mu'_1, \mu'_2)$ | $A$ and $B$ respective estimations are not close which leads to error detection and finally discarding. | $A$ and $B$ respective estimations are not close which leads to error detection and finally discarding. |

This is a heuristic reasoning, and we must rather consider most general strategies $\omega(i, j, \mu_1, \mu_2, \mu'_1, \mu'_2)$ and write the probability equations with the appropriate group transform, under Deep Random assumption, which is done in [9]. But this little array explains why we create partial independence between the BSC and consequently then an advantage for the legitimate partners compared to the opponent, bearing in mind that $(\mu_1, \mu_2)$ (resp. $(\mu'_1, \mu'_2)$) are absolutely undistinguishable knowing $i$ (resp. $j$), due to the fact that the distributions $\Phi$ and $\Psi$ (resp. $\Phi'$ and $\Psi'$) are unknown and thus also absolutely undistinguishable by $\xi$.

The sampling method presented above has the drawback of the border effect. If the reference value $V_B$ is too close from one of the sampling frontier $\left\{\frac{tK}{\sqrt{nk}}\right\}_{t\in\mathbb{N}}$, then the sampling process becomes unefficient. In order to avoid the border effect, one can bring a little improvement to the protocol by allowing $B$ to publish :

$$\rho_B = V_B - \frac{K}{\sqrt{nk}}\left\lfloor V_B \frac{\sqrt{nk}}{K}\right\rfloor$$

and then to replace $V_B$ by $V_B \to V_B + \frac{K}{2\sqrt{nk}} - \rho_B$ in order to center $V_B$ within the sampling comb. This of course results in applying the same transform on $V_A$ and $\omega$ :

$$V_A = \frac{K}{\sqrt{nk}}\left\lfloor\left(V_A + \frac{K}{2\sqrt{nk}} - \rho_B\right)\frac{\sqrt{nk}}{K}\right\rfloor + \frac{K}{2\sqrt{nk}}$$

$$\omega = \frac{K}{\sqrt{nk}}\left\lfloor\left(\omega + \frac{K}{2\sqrt{nk}} - \rho_B\right)\frac{\sqrt{nk}}{K}\right\rfloor + \frac{K}{2\sqrt{nk}}$$

The publication of $\rho_B$ does not bring any additional information to the opponent regarding the valuable secret information being the parity of $\left\lfloor V_B \frac{\sqrt{nk}}{K} \right\rfloor$.

The reception of $\widetilde{e_A}$ as $\widetilde{e_B}$ by $B$, and as $\widetilde{e_\xi}$ by $\xi$ corresponds to a broadcast BSC channel (we can easily see that the channels are BSC by remarking that changing $\rho$ in $\rho + \frac{K}{\sqrt{nk}}$ inverses the digits and is independent of $V_A$, $V_B$, and $V_\xi$). That BSC can be modelized in a $BSC(\epsilon_A, \epsilon_B, \epsilon_\xi)$ ([9] Proposition 13) where the 3 independent BSC have binary error variables $(\delta_A, \delta_B, \delta_\xi)$ with respective binary error probability $(\epsilon_A, \epsilon_B, \epsilon_\xi)$. It has been shown by Maurer in [5] that in such model, if $\epsilon_\xi > 0$ then Advantage Distillation can be achieved thanks to error correcting methods, and consecutively perfect secrecy can be approached as close as desired thanks to Information Reconciliation and Privacy Amplification techniques. One can then use the non-optimal error correcting method described by Maurer in [5] to reconcile digit flow between the legitimate partner: the codeword $v_A$ chosen by $A$ can only be $(0,0,\dots,0)_L$ or $(1,1,\dots,1)_L$ depending on $e_A = 0$ or $e_A = 1$. $B$ publicly discards all decoded sequence $v_B$ that is not $(0,0,\dots,0)_L$ or $(1,1,\dots,1)_L$ and obviously decodes accordingly $e_B = 0$ if $|v_B| = 0$, and $e_B = 1$ if $|v_B| = L$.

## II. Extended protocol against active opponent

The protocol $\mathcal{P}$ introduced in [9] and summarized in the former section is suitable to generate a common secret bit string $S$ between the legitimate partners against a passive unlimited opponent. The parameters of $\mathcal{P}$ can be adapted so that, $\forall \varepsilon, \varepsilon' > 0$

*(i)* $\quad P(e_A \neq e_B) \leq \varepsilon$

*(ii)* $\quad \left| P(e_\xi = e_B) - \frac{1}{2} \right| \leq \varepsilon'$

In order to generate $S$, the partners will execute a sequence of $L$ rounds of $\mathcal{P}$ that can be serialized as follows:

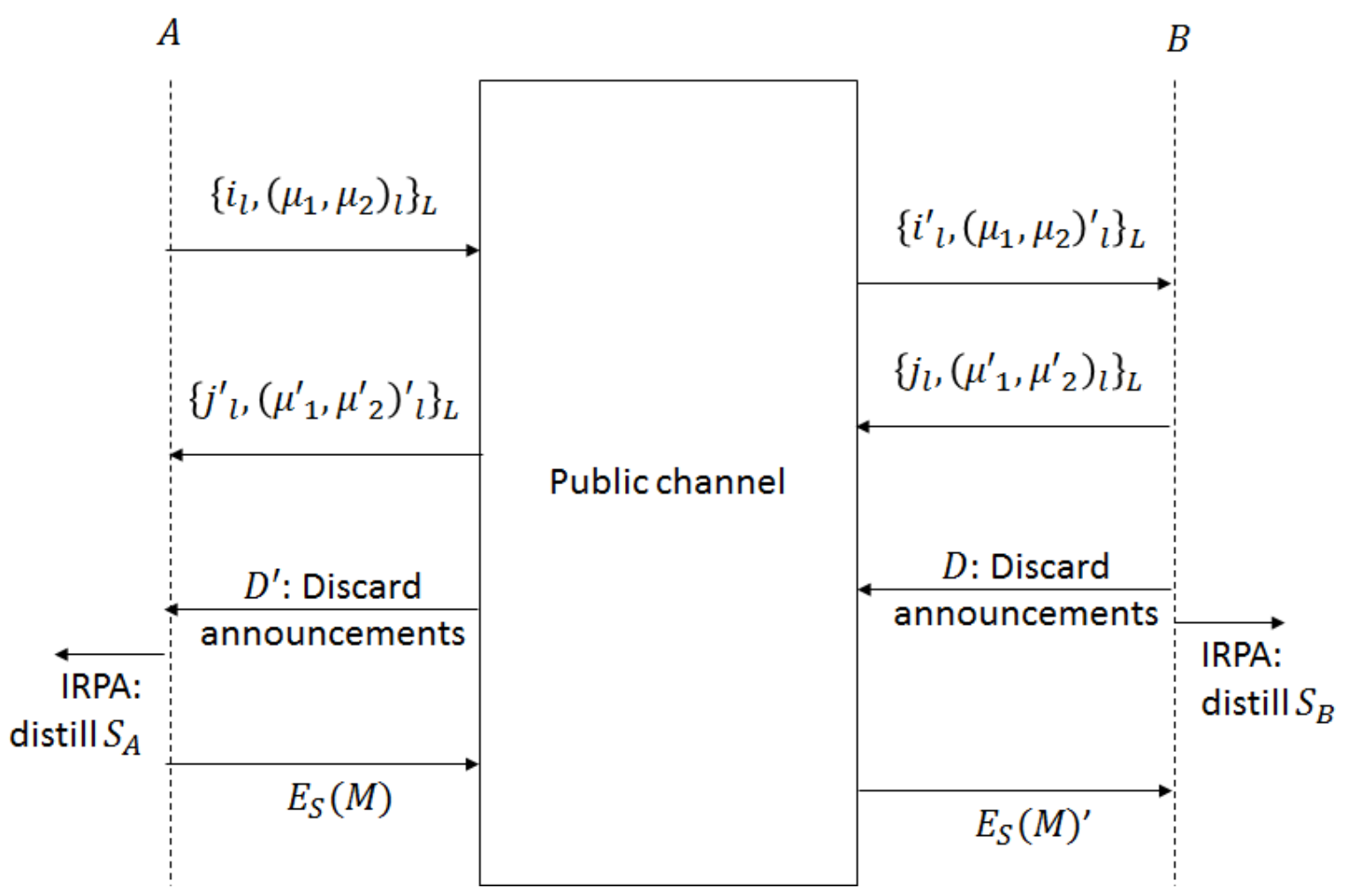

The purpose of generating $S$ is of course to transmit a secret message $M$ from $A$ to $B$. The encryption function $E_S(M)$ can be simply $S \oplus M$ if one want the highest security offered by one time pad, or another symmetric key encryption scheme working with a shared secret key $S$. While we have shown that a passive opponent cannot gain knowledge of $S$, it is obvious that an active opponent can insert himself in the communication, first playing the role of $B$ with $A$, which enables him to collect $M$, and then, optionally, to play the role of $A$ with $B$ to transmit $M$, or even a different message $M'$ to $B$.

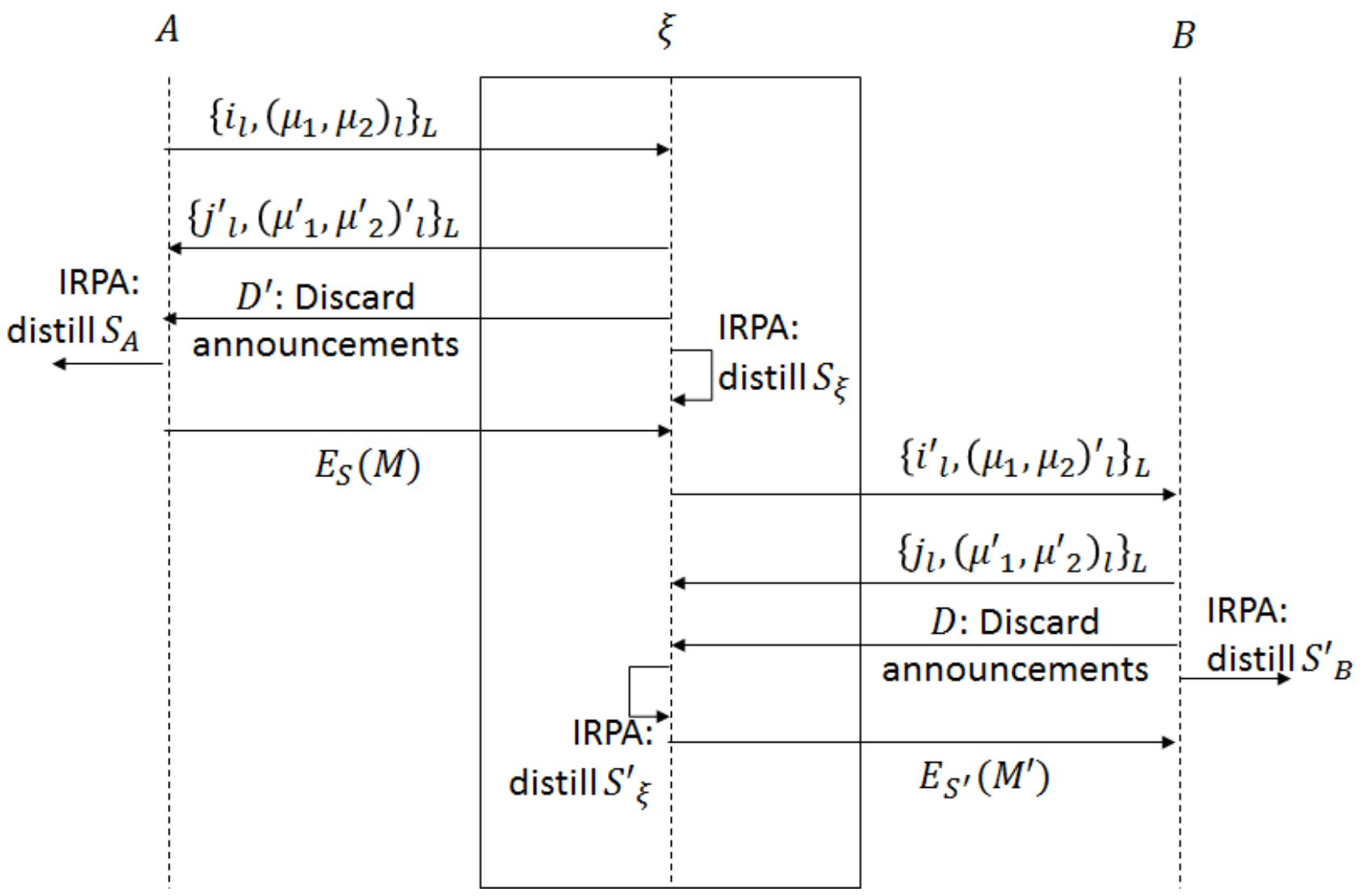

We will see now how to prevent this MITM attack with an extended version of the protocol $\mathcal{P}$.

**The security model**

In our 'active opponent scenario', $\xi$ still has unlimited computation and storage power, but is also capable to (i) delete messages from the public channel, (ii) introduce new messages in the public channel at destination of either $A$ or $B$; (i)+(ii) is also equivalent to the capacity to modify a message transiting from $A$ to $B$ or from $B$ to $A$.

$A$ and $B$ are equipped with a DRG, and also with a private 'wallet' that can contain a shared authentication secret $s$, that can be updated at any time by the wallet holder. It is assumed as a pre-condition to the protocol that $A$ and $B$ have initially the same value $s_0$ in their wallet.

It is assumed that the opponent $\xi$ has no access to (i) the content of a private wallet, (ii) anything computed or stored within a DRG.

The goal of the extended protocol is to continue to ensure that, $\forall \varepsilon, \varepsilon' > 0$

*(i)* $\quad P(e_A \neq e_B) \leq \varepsilon$

*(ii)* $\quad \left|P(e_\xi = e_B) - \frac{1}{2}\right| \leq \varepsilon'$

even with such an active opponent.

**Description of the extended protocol $\mathcal{P}^*$**

The extended protocol that we propose is possible only because of a specific non-reversibility property of $\mathcal{P}$. Indeed it is shown in [9] (Lemma3) that, when the public information $(i, j)$ is given, with the assumption that the distributions are synchronized, then it is impossible to reversely determine $x$ or $y$ with an accuracy equal to the one of a partner's estimation. $\mathcal{P}^*$ is original because, unlike any other protocol that does not benefit from the non-reversibility property of Deep Random Secrecy, it resists to opponents that are both unlimited and active MITM.

The extension then consists in adding a mutual verification phase after the bit string $S$ has been generated, and before transmitting the secret message $M$. This verification is performed by:

(i) dividing the shared secret into 2 independent piece $(s_A, s_B)$ (with $H(s_A) = H(s_B) = H(s)/2$)

(ii) sending from $A$ to $B$ (resp. $B$ to $A$) a verification code

$$c_A = h(s_A, \{i_l, (\mu_1, \mu_2)_l\}_L, \{j'_l, (\mu'_1, \mu'_2)'_l\}_L)$$

(resp.)

$$c_B = h(s_B, \{i'_l, (\mu_1, \mu_2)'_l\}_L, \{j_l, (\mu'_1, \mu'_2)_l\}_L)$$

(the characteristics of the verification function $h$ will be discussed hereafter)

(iii) verifying by $B$ (resp. by $A$) that the local computation

$$v_A = h(s_A, \{i'_l, (\mu_1, \mu_2)'_l\}_L, \{j_l, (\mu'_1, \mu'_2)_l\}_L)$$

(resp.)

$$v_B = h(s_B, \{i_l, (\mu_1, \mu_2)_l\}_L, \{j'_l, (\mu'_1, \mu'_2)'_l\}_L)$$

is equal to the received code $(c'_A = v_A)$ (resp. $(c'_B = v_B)$

(iv) using part of the entropy of $S$ to renew (sending $R(s')$) a new value of the shared authentication secret $s'$ for a next round

(v) storing by $A$ (resp. $B$) the new value of shared authentication secret $s'$ in its private wallet.

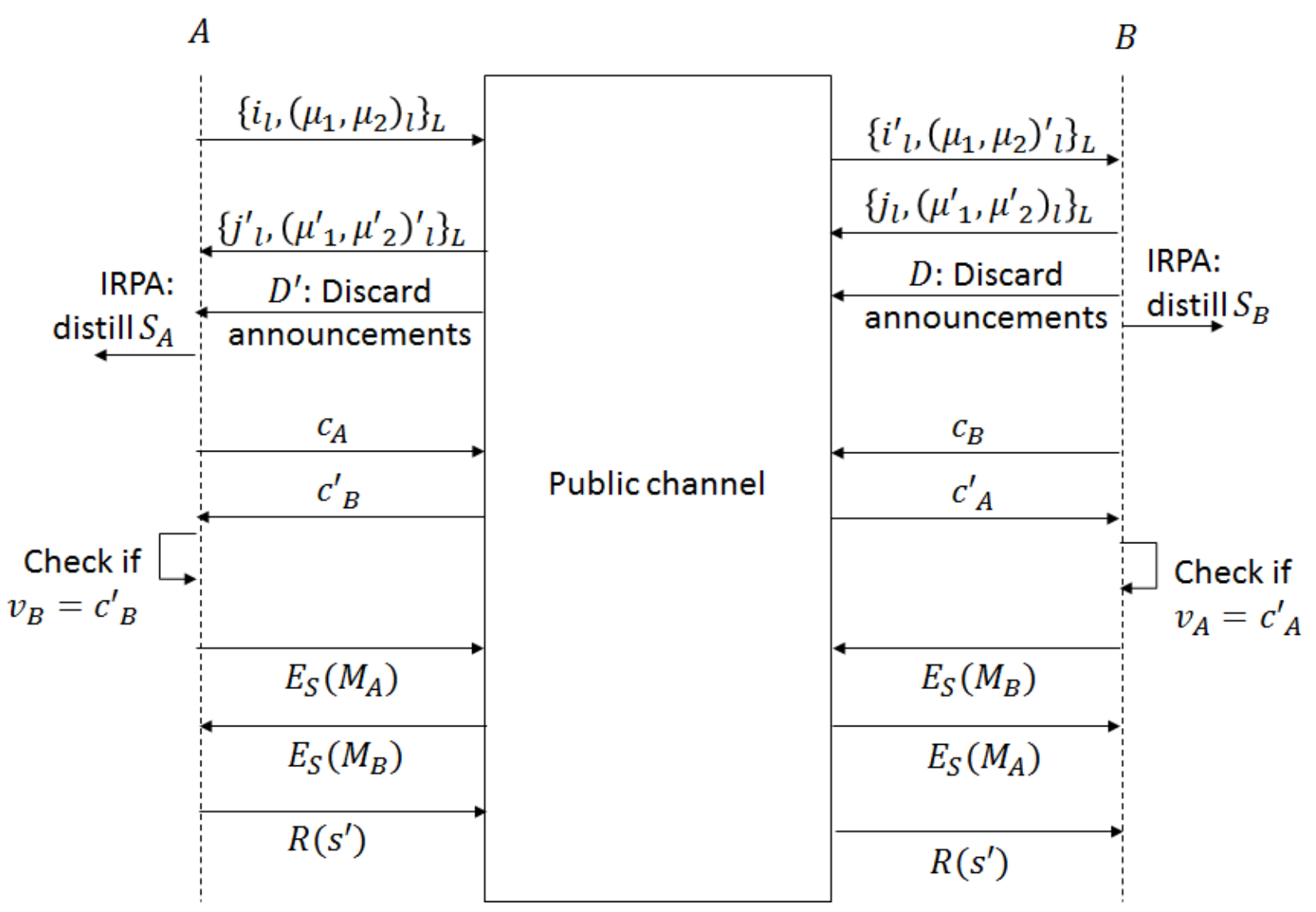

Assuming that $E_S(M) = S \oplus M$, $M$ having a length bounded at each round by $|M| \leq L_M$, the common shared bit string must have length:

$$|S| \geq H(s) + L_M$$

in order to be able to fully renew the shared authentication secret $s$ after each round.

**Design considerations for security**

The verification function $h$ must have at least the following characteristics:

Property 1:

$$if\ \{i_l, (\mu_1, \mu_2)_l\}_L, \{j'_l, (\mu'_1, \mu'_2)'_l\}_L\ are\ known,\ then$$

$$\#\{s | h(s, \{i_l, (\mu_1, \mu_2)_l\}_L, \{j'_l, (\mu'_1, \mu'_2)'_l\}_L) = h(s_A, \{i_l, (\mu_1, \mu_2)_l\}_L, \{j'_l, (\mu'_1, \mu'_2)'_l\}_L)\} \gg 1$$

Property 2:

$$if\ \{i_l, (\mu_1, \mu_2)_l\}_L, \{j'_l, (\mu'_1, \mu'_2)'_l\}_L\ are\ known,\ and\ s \neq s'$$

$$P\big(h(s, \{i_l, (\mu_1, \mu_2)_l\}_L, \{j'_l, (\mu'_1, \mu'_2)'_l\}_L) = h(s', \{i_l, (\mu_1, \mu_2)_l\}_L, \{j'_l, (\mu'_1, \mu'_2)'_l\}_L)\big) = o_{H(s)}(1)$$

Due to the non-reversibility property of $\mathcal{P}$, it is not possible for $\xi$ to keep $i_l, (\mu_1, \mu_2)_l$ (resp. $j_l, (\mu'_1, \mu'_2)_l$) unchanged when transmitting towards $B$ (resp. $A$) because then it is impossible to generate $x$ and $\sigma_\Phi$ (resp. $y$ and $\sigma_{\Phi'}$) that would enable $\xi$ to play the role of $A$ vis à vis $B$ (resp. $B$ vis à vis $A$) in determining $S_\xi = S_A = S_B$ with unchanged transmitted public information $i_l, (\mu_1, \mu_2)_l$ and $j_l, (\mu'_1, \mu'_2)_l$. Therefore, $\xi$ has no other choice than to execute fairly $\mathcal{P}$ with both $A$ and $B$ in order to share a secret with either of them. $\xi$ can try to adjust the value $s_\xi$ of a guessed shared authentication secret, but due to the fact that there exists many possible values of $s_\xi$ such that

$$h\big(s_{\xi,A}, \{i_l, (\mu_1, \mu_2)_l\}_L, \{j'_l, (\mu'_1, \mu'_2)'_l\}_L\big) = h(s_A, \{i_l, (\mu_1, \mu_2)_l\}_L, \{j'_l, (\mu'_1, \mu'_2)'_l\}_L)$$

and

$$h\big(s_{\xi,B}, \{i'_l, (\mu_1, \mu_2)'_l\}_L, \{j_l, (\mu'_1, \mu'_2)_l\}_L\big) = h(s_B, \{i'_l, (\mu_1, \mu_2)'_l\}_L, \{j_l, (\mu'_1, \mu'_2)_l\}_L)$$

it is not possible for $\xi$ to guarantee that

$$h\big(s_{\xi,A}, \{i_l, (\mu_1, \mu_2)_l\}_L, \{j'_l, (\mu'_1, \mu'_2)'_l\}_L\big) = h(s_A, \{i'_l, (\mu_1, \mu_2)'_l\}_L, \{j_l, (\mu'_1, \mu'_2)_l\}_L)$$

and

$$h\big(s_{\xi,B}, \{i'_l, (\mu_1, \mu_2)'_l\}_L, \{j_l, (\mu'_1, \mu'_2)_l\}_L\big) = h(s_B, \{i_l, (\mu_1, \mu_2)_l\}_L, \{j'_l, (\mu'_1, \mu'_2)'_l\}_L)$$

for all $l$.

One may question if it is possible for $\xi$ to learn at least some bits of $S$. If $h$ is not carefully designed, here is how $\xi$ can learn even almost all bits of $S$. Assume for example that we choose:

$$h(s, \{i_l, (\mu_1, \mu_2)_l\}_L, \{j'_l, (\mu'_1, \mu'_2)'_l\}_L) = f\left(s, \left(\bigoplus_{l=1}^{L} (i_l \oplus j'_l)\right), \left(\bigoplus_{l=1}^{L} ((\mu_1, \mu_2)_l \oplus (\mu'_1, \mu'_2)'_l)\right)\right)$$

$\xi$ can then sacrifice the last instance $j'_L, (\mu'_1, \mu'_2)'_L$ to control that $c'_B = v_B$ without knowing $s_B$, by resolving the equations:

$$\left(\bigoplus_{l=1}^{L} i_l\right) \oplus \left(\bigoplus_{l=1}^{L-1} j'_l\right) \oplus j'_L = \left(\bigoplus_{l=1}^{L} i'_l\right) \oplus \left(\bigoplus_{l=1}^{L} j_l\right)$$

$$\left(\bigoplus_{l=1}^{L} (\mu_1, \mu_2)_l\right) \oplus \left(\bigoplus_{l=1}^{L-1} (\mu'_1, \mu'_2)'_l\right) \oplus (\mu'_1, \mu'_2)'_L = \left(\bigoplus_{l=1}^{L} (\mu_1, \mu_2)'_l\right) \oplus \left(\bigoplus_{l=1}^{L} (\mu'_1, \mu'_2)_l\right)$$

The interaction flows perform as follows:

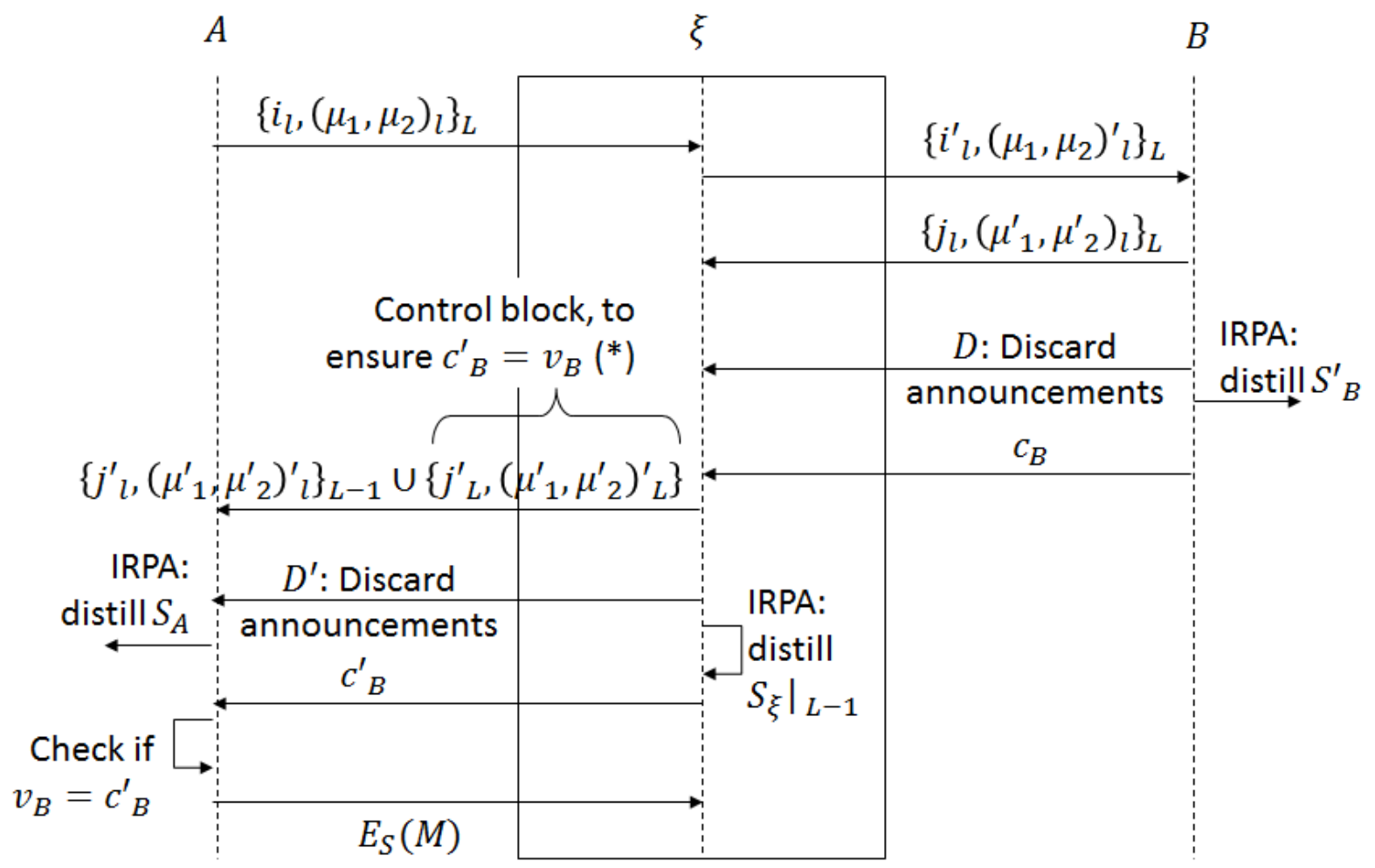

By proceeding as described above, $\xi$ can gain the amount of knowledge of $S$ given by performing correctly all instances except the last one, which will give

$$I(S_\xi; S) \geq \frac{L-1}{L} H(S)$$

This attack is made possible because of the existence of transformation over $\{j'_l, (\mu'_1, \mu'_2)'_l\}_L$ and/or $\{i'_l, (\mu_1, \mu_2)'_l\}_L$ that can let $h$ invariant independently of $s$. The design of $h$ should then guarantee:

Property 3:

> *There does not exist any non-trivial transformation $T$ applying on the space $\{i'_l, (\mu_1, \mu_2)'_l, j'_l, (\mu'_1, \mu'_2)'_l\}_L$ such that*
>
> $$\forall s, \quad h(s, \{i_l, (\mu_1, \mu_2)_l\}_L, T(\{j'_l, (\mu'_1, \mu'_2)'_l\}_L)) = h(s, T(\{i'_l, (\mu_1, \mu_2)'_l\}_L), \{j_l, (\mu'_1, \mu'_2)_l\}_L)$$
>
> or
>
> $$\forall s, \quad h(s, T(\{i'_l, (\mu_1, \mu_2)'_l\}_L), \{j_l, (\mu'_1, \mu'_2)_l\}_L) = h(s, \{i_l, (\mu_1, \mu_2)_l\}_L, T(\{j'_l, (\mu'_1, \mu'_2)'_l\}_L))$$

We propose to use:

$$h_{a,b}(x) \triangleq \big((ax+b) \bmod p\big) \bmod 2^{\frac{H(S_A)}{2}}$$

$$\begin{aligned} h(s, \{i_l, (\mu_1, \mu_2)_l\}_L, \{j_l, (\mu'_1, \mu'_2)_l\}_L) \\ \triangleq \Big(\oplus_{l=1}^{L} \big(h_{a_l,b_l}(i_l + s\overline{\iota_l})\big)\Big) \oplus \Big(\oplus_{l=1}^{L} \big(h_{a'_l,b'_l}(j_l + s\overline{\jmath_l})\big)\Big) \\ \oplus \Big(\oplus_{l=1}^{L} \big(h_{c_l,d_l}(\mu_{1_l} + s\overline{\mu_{1_l}})\big)\Big) \oplus \Big(\oplus_{l=1}^{L} \big(h_{c'_l,d'_l}(\mu_1{}'_l + s\overline{\mu_1{}'_l})\big)\Big) \\ \oplus \Big(\oplus_{l=1}^{L} \big(h_{e_l,f_l}(\mu_{2_l} + s\overline{\mu_{2_l}})\big)\Big) \oplus \Big(\oplus_{l=1}^{L} \big(h_{e'_l,f'_l}(\mu_2{}'_l + s\overline{\mu_2{}'_l})\big)\Big) \end{aligned}$$

where $p$ is a prime superior to $2^{\frac{H(S)}{2}}$, and $\{a_l, b_l, c_l, d_l, e_l, f_l, a'_l, b'_l, c'_l, d'_l, e'_l, f'_l\}$ are random non-zero integers, fixed as parameters of the protocol (not negotiated); they can typically be the successive outputs of a congruential generator to avoid to store all them in memory. The permutation $\mu_{1_l}$ is written above with its canonical binary form to be able to compute it with arithmetic. $i_l + s\overline{\iota_l}$ is a non-linear form on $i_l$ also involving $s$ for each $i_l, \mu_{1l}, \mu_{2l}, j_l, \mu'_{1l}, \mu'_{2l}$. $\{h_{a,b}(x)\}$ is a well known universal hashing class of functions, thus benefiting from the good probabilistic properties of such class.

The modularity structure of $h$ in terms of using blocks of the form $\left(h_{a_l,b_l}(i_l + s\overline{\iota_l})\right)$ and composing them by simple $\oplus$ is chosen on purpose; in case public information $\{i_l, (\mu_1, \mu_2)_l\}$ are reused from a round to another, the associated block can be reused without re-computation of the arithmetical operations. This reuse technique will be illustrated in optimization techniques for $\mathcal{P}$, presented in further work.

**Other considerations**

Eventually, note that the computation of $h$ is obviously highly sensitive to errors in the transmission of the $\{i_l, (\mu_1, \mu_2)_l, j_l, (\mu'_1, \mu'_2)_l\}_L$, and therefore it is recommended to associate each $\{i_l, (\mu_1, \mu_2)_l\}$ with an error control checksum.

The size parameters $H(s), L_M, L$ should also be fixed as parameters of the protocol and not negotiated, in order to avoid that they become also vulnerable to MITM attacks.

## III. Conclusion

We have proposed an authentication scheme resisting to unlimited MITM attackers. This scheme is working in conjunction with a key exchange protocol and is only made possible by the very special non-reversibility property of Deep Random Secrecy. The scheme requires a prior shared authentication secret that is kept fully independent from the shared encryption key generated between the partners as the output of the protocol. The shared authentication secret has to be renewed after each round of the protocol for a full secrecy against unlimited opponent, and the renewal process is part of the protocol.

This protocol is then suitable to replace protocols like TLS in scenarios where opponents are assumed unlimited (typically because equipped with quantum computers), but requires a prior registration phase between the partners to initialize the shared authentication secret. Such registration is quite realistic in practice (think about creating an online account for instance). However, one could question the poor level of performance (bandwidth, CPU) of the protocol to replace a popular protocol like TLS.

The main performance concern is of course the quantity of information needed to generate an output shared key $S$. That quantity is superior to $|S|Rn\log n$ (where $R$ is a constant depending on IRPA method used in $\mathcal{P}^*$). In general, such a key exchange protocol based on Deep Random Secrecy and resistant to unlimited MITM cannot exceed a bandwidth performance of $|S|/C$, where $C \geq 1$ is the Cryptologic Limit introduced in [9]. The question of optimization of $\mathcal{P}$ (and $\mathcal{P}^*$) in terms of getting close to $C$ will be addressed in further work.

The CPU performance question can be addressed by searching fastest possible functions $h$ that verifies properties 1, 2 and 3. It can also be questioned if one can lower the requirement of Property 3 together with maintaining the security model; Property seems too much demanding at first sight compared to the objective because many transformations $T$ could be harmless to the protocol.

**Who is the author ?**

I have been an engineer in computer science for 20 years. My professional activities in private sector are related to IT Security and digital trust, but have no relation with my personal research activity in the domain of cryptology. If you are interested in the topics introduced in this article, please feel free to establish first contact at tdevalroger@gmail.com